\documentclass[twocolumn,aps,prb]{revtex4}

\usepackage{dcolumn}
\usepackage{bm}
\usepackage{latexsym,amssymb,float}
\usepackage{setspace}
\usepackage{graphicx}
\usepackage{epsfig}

\usepackage{amsmath}
\usepackage{bm}

\usepackage[english]{babel}
\usepackage{verbatim}

\begin{document}

\title{Defects in Heavy-Fermion Materials: Unveiling Strong Correlations in Real Space}
\author{Jeremy Figgins and Dirk K. Morr}
\affiliation{Department of Physics, University of Illinois at
Chicago, Chicago, IL 60607, USA}
\date{\today}

\begin{abstract}
{\bf Complexity in materials often arises from competing
interactions at the atomic length scale. One such example are the
strongly correlated heavy-fermion materials where the competition
\cite{Don77} between Kondo screening \cite{Kondo64} and
antiferromagnetic ordering  is believed to be the origin of their
puzzling non-Fermi-liquid properties
\cite{Map95,ASch00,Ste01,Cus03,Loh07,Geg08}. Insight into such
complex physical behavior in strongly correlated electron systems
can be gained by impurity doping \cite{Bal06,Yaz99,Hud99,Hud01}.
Here, we develop a microscopic theoretical framework to
demonstrate that defects implanted in heavy-fermion materials
provide an opportunity for unveiling competing interactions and
their correlations in real space. Defect-induced perturbations in
the electronic and magnetic correlations possess
characteristically different spatial patterns that can be
visualized via their spectroscopic signatures in the local density
of states or non-local spin susceptibility. These real space
patterns provide insight into the complex electronic structure of
heavy-fermion materials, the light or heavy character of the
perturbed states, and the hybridization between them. The strongly
correlated nature of these materials also manifests itself in
highly non-linear quantum interference effects between defects
that can drive the system through a first-order phase transition
to a novel inhomogeneous ground state.}

\end{abstract}

\maketitle

The essential features of heavy fermion materials is a Kondo
lattice of magnetic atoms with localized moments arising from $4f$
or $5f$-electron shells, and a delocalized conduction band
\cite{Hew93}. The magnetic moments are coupled to each other via
an antiferromagnetic interaction, $I>0$, and to the spin moments
of the conduction electrons via the Kondo coupling, $J>0$
\cite{Don77,Kondo64}. The ground state of this system is realized
either by Kondo screening of the magnetic moments, resulting in a
heavy-Fermi-liquid phase \cite{Hew93}, or by their
antiferromagnetic ordering. These two competing phases, which
manifest strongly correlated many-body states, are separated by a
quantum critical point (QCP). To date, no theoretical consensus
has been reached \cite{Ful88,Col01,Si01,Sun03,Sen03,Paul07} on the
physical origin for the astonishing deviations from Landau's
Fermi-liquid theory observed in the quantum critical region above
the QCP \cite{Map95,ASch00,Ste01,Cus03,Loh07,Geg08}. Understanding
how the formation of antiferromagnetism competes locally with the
creation of a Kondo singlet, how coherence is established in Kondo
lattices, and how the resulting magnetic and electronic
correlations are intertwined, is crucial in unraveling this
mystery.

By their very nature defects implanted in a Kondo lattice
\cite{Ste87,Lin87,Lop92,Law96} manifest local perturbations that
can reveal the form of strong correlations in real space. These
defects can take the form of missing magnetic Kondo atoms, i.e.,
Kondo holes, or of non-magnetic impurities. By differently
perturbing the electronic and magnetic states of the system,
defects provide an unprecedented opportunity to differentiate (in
real space) between electronic correlations arising from Kondo
screening, and the antiferromagnetic correlations between the
magnetic moments, as schematically shown in Fig.~S1 of the
Supplementary Information. The spatial form and extent of these
perturbations are not only a measure for the strength of the
correlations, but also provide insight into the hybridization of
the (heavy) magnetic $f$-electron states with the (light)
conduction band. This hybridization results in the Kondo screening
of the magnetic moments and the opening of a gap in the electronic
spectrum. The experimentally accessible, spectroscopic
fingerprints of the perturbed electronic and magnetic correlations
can be found in a redistribution of spectral weight in the local
density of states of the conduction band, and in an enhancement or
suppression of the $f$-electron spin susceptibility, respectively.
Modulations in the density of the conduction electrons, as induced
by a non-magnetic impurity, have profound effects on the spatial
form of the correlations, and lead to the formation of an impurity
bound inside the hybridization gap. The coupling between
electronic and magnetic correlations can create highly non-linear
feedback effects on quantum interference in periodic arrays of
Kondo holes, leading to the spatial reconstruction of the ground
state via a first order phase transition. Our findings open a new
path for studying competing interactions in real space, and
provide a microscopic, real space picture of the perturbations
induced by defects in heavy fermion materials.

In order to study the spatial perturbations induced by defects in
the magnetic and electronic correlations of a (heavy fermion)
Kondo lattice system, we have developed a real-space formalism,
the {\it Kondo-Bogoliubov-de Gennes method}, whose details are
described in Sec.~S2 of the Supplementary Information. This
theoretical formalism provides us with two salient quantities that
determine the physical properties of the system. The first one is
the site-dependent hybridization, $s({\bf r})$, between the
conduction band and the $f$-electron state of a Kondo atom. It is
a measure of the Kondo screening of the magnetic moment, and of
the electronic correlations in the resulting many-body state.
Here, we focus on the Kondo screened phase of the heavy-fermion
materials where $s({\bf r}) \not = 0$ for all sites. The second
quantity, $\chi({\bf r},{\bf r}')$, represents the strength of
short-range antiferromagnetic correlations between magnetic
moments on (nearest-neighbor) sites ${\bf r}$ and ${\bf r}'$
\cite{Sen03,Paul07}.

\begin{figure}[t]
\includegraphics[width=8.5cm]{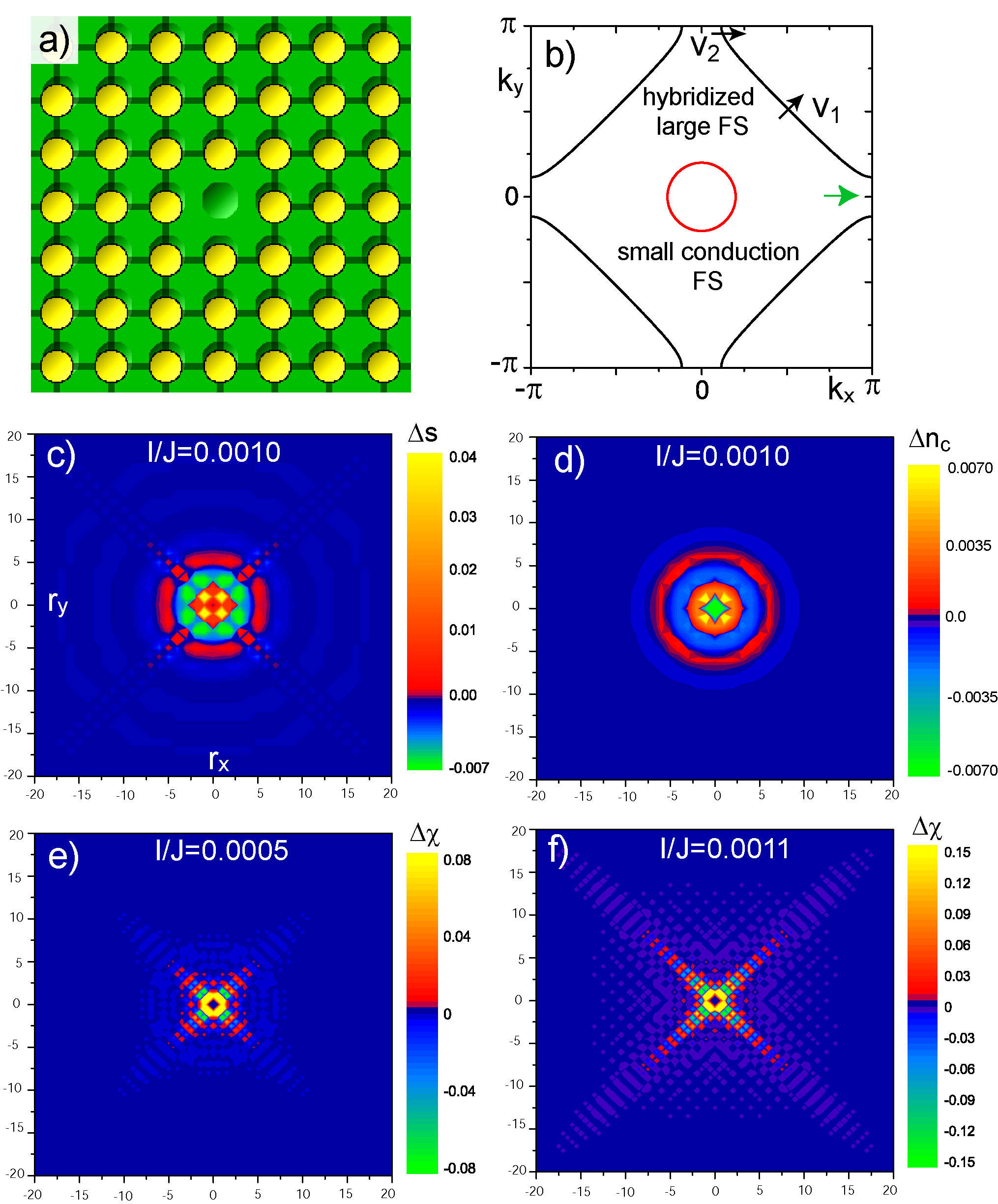}
\caption{{\bf A hole in the Kondo lattice} {\bf a.} Schematic
picture of a Kondo lattice containing a Kondo hole: green (orange)
dots represent the sites of the conduction lattice (magnetic
atoms). {\bf b.} Electronic Structure of an unperturbed Kondo
lattice system for $I/J=0.001$: the large Fermi surface (black)
arises from the hybridization of the magnetic $f$-electron states
with the conduction band, the small Fermi surface (red) represents
the (unhybridized) conduction band. Spatial contour plots of {\bf
c.} $\Delta s$ and {\bf d.} $\Delta n_c$ for $I/J=0.001$. Spatial
contour plots of $\Delta \chi$ for {\bf e.} $I/J=0.0005$, and {\bf
f.} $I/J=0.0011$. The relative changes, $\Delta s$, $\Delta n_c$,
and $\Delta \chi$, are obtained by subtracting the results for
$s({\bf r})$, $n_c({\bf r})$ and $\chi({\bf r},{\bf r}')$ in the
unperturbed Kondo lattice from those of the Kondo hole system, and
dividing by the former. We display the value of $\chi({\bf r},{\bf
r}')$ between nearest neighbor sites ${\bf r}$ and ${\bf r}'$ at
$({\bf r}+{\bf r}')/2$.} \label{fig:Fig1}
\end{figure}
The simplest defect in a Kondo lattice is realized by removing a
magnetic Kondo atom from it, resulting in a Kondo hole, as shown
in Fig.~\ref{fig:Fig1}a. To demonstrate the spatial perturbations
in the electronic and magnetic correlations of the system that are
induced by the Kondo hole, we present in Figs.~\ref{fig:Fig1}c and
\ref{fig:Fig1}d, a two-dimensional contour plot of the relative
change in the hybridization, $\Delta s$, and the conduction
electron density, $\Delta n_c$, respectively, between the Kondo
lattice with and without a hole. Both quantities exhibit similar
spatial oscillations, whose isotropy and wavelength of $5a_0$
($a_0$ being the lattice constant) imply that they are determined
by the Fermi surface of the unhybridized conduction band (see
Fig.~\ref{fig:Fig1}b) with Fermi wavelength $\lambda^c_F=10 a_0$.
These oscillations decay exponentially but remain practically
unchanged over the range of $I/J$ considered here. The spatial
plots of $\Delta \chi$ in Figs.~\ref{fig:Fig1}e and
\ref{fig:Fig1}f reveal strongly anisotropic oscillations along the
lattice diagonal. Their weaker reflection can also be found in
$\Delta s$ (see Fig.~\ref{fig:Fig1}c), clearly demonstrating the
coupling between the system's electronic and magnetic
correlations. The spatial pattern of $\Delta \chi$ is determined
by the strongly anisotropic Fermi surface of the hybridized
system, shown in Fig.~\ref{fig:Fig1}b, which possesses a large
degree of nesting and a Fermi velocity, $v_1$, along the lattice
diagonal which is about 10 times larger than that along the bond
direction, $v_2$. This conclusion receives further support from
the rapid oscillations in $\Delta \chi$ along the lattice diagonal
(see Fig.~S2 in the Supplementary Information) which possess a
wavelength of $\lambda^f_F/2= \sqrt{2} a_0$, with $\lambda^f_F$
being the Fermi wavelength of the hybridized Fermi surface along
the diagonal. The envelope of these oscillations decays
exponentially with distance, $r$, from the Kondo hole, i.e.,
$\Delta \chi \propto \exp{\left(-r/\xi \right)}$. Such an
exponential decay is expected since within the KBdG formalism, a
Kondo hole is equivalent to an $f$-electron state with on-site
energy $\varepsilon_f = \infty$. Such a state it localized, and
its effects on $\Delta \chi$ or $\Delta s$ necessarily decay
exponentially. With increasing magnetic interaction, $\xi$
increases approximately linearly  (see Fig.~S2 in the
Supplementary Information), leading to a the growth in the spatial
extent and amplitude of the perturbations in $\Delta \chi$, as can
directly be seen from a comparison of Figs.~\ref{fig:Fig1}e and
\ref{fig:Fig1}f. These perturbations are therefore a direct
measure for the strength of the magnetic interaction and the
resulting correlations in the material.

\begin{figure}[t]
\includegraphics[width=8.5cm]{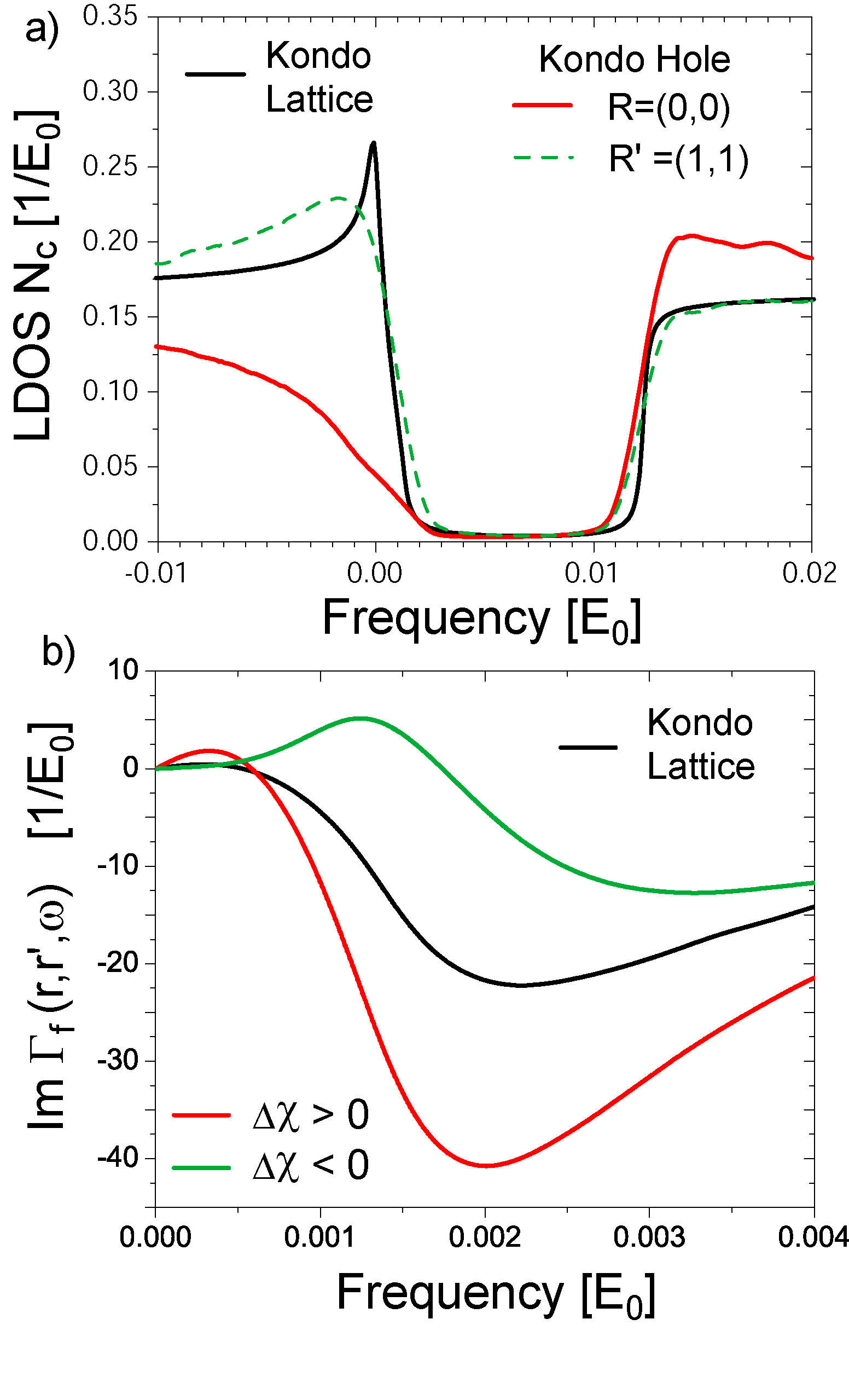}
\caption{{\bf Spectroscopic Fingerprints} {\bf a.} The local
conduction band density of state $N_c({\bf r},\omega)$ for the
unperturbed Kondo lattice, and at the site of the Kondo hole,
${\bf R}=(0,0)$, and its next-nearest-neighbor site ${\bf
R}'=(1,1)$ where $\Delta s({\bf R}')>0$. {\bf b.} The non-local
$f$-electron spin susceptibility, $\Gamma_f ({\bf r, r'}, \omega)$
between nearest neighbor sites in the unperturbed Kondo lattice,
and in the Kondo hole system between sites ${\bf r} = (1,1)$ and
${\bf r'}=(1,0)$ with $\Delta \chi >0$ and between sites ${\bf r}
= (1,1)$ and ${\bf r'}=(1,2)$ with $\Delta \chi <0$.}
\label{fig:spec}
\end{figure}
The spatial perturbations in the electronic correlations, $s({\bf
 r})$ possess a direct spectroscopic signature in the local density
of states of the conduction band, $N_c({\bf r},\omega)$, which can
be probed via scanning tunneling spectroscopy (STS)
\cite{Li98,Ujs00,Mad01,Mal09,Fig09}. The LDOS describes the
quantum mechanical probability to add (for $\omega>0$) or remove
(for $\omega<0$) a conduction electron with energy $\hbar \omega$
from the system at site ${\bf r}$. $N_c({\bf r},\omega)$ for an
unperturbed Kondo lattice, shown in Fig.~\ref{fig:spec}a, exhibits
a gap which is the primary spectroscopic signature of the
hybridization between the conduction band and $f$-electron state
of the magnetic atoms, and the concomitant screening of the
magnetic moments. This gap was recently observed for the first
time in STS experiments by Schmidt {\it et al.}~\cite{Sch09}. The
peak in the LDOS at the low-energy site of the gap arises from the
Van Hove singularity of the hybridized Fermi sea (indicated by the
green arrow in Fig.~\ref{fig:Fig1}b). A comparison with the LDOS
at the site of the Kondo hole reveals a significant shift of
spectral weight from negative to positive energies. To understand
this redistribution, we note that the screening of a single Kondo
atom by conduction electrons leads to an increase in the local
electron density, $n_c({\bf r})$, which is related to the LDOS via
\begin{equation}
n_c({\bf r})=\int_{-\infty}^{\infty} d\omega \; n_F(\omega) \;
N_c({\bf r},\omega)
\end{equation}
with $n_F$ being the Fermi distribution function. An increase in
$n_c({\bf r})$ therefore necessarily implies a redistribution of
spectral weight in $N_c({\bf r},\omega)$ from positive to negative
energies. The removal of a Kondo atom from the lattice simply
leads to the opposite effect with a decrease in $n_c({\bf r})$ at
the site of the hole, and the corresponding shift in the LDOS
shown in Fig.~\ref{fig:spec}a. This suggests that $N_c({\bf
r},\omega)$ directly reflects the perturbations in $s({\bf r})$,
since for a site with $\Delta s({\bf r})>0$ one has $\Delta
n_c({\bf r})>0$, and a concomitant redistribution of spectral
weight in $N_c({\bf r},\omega)$ to negative frequencies (and vice
versa). Indeed, a plot of $N_c({\bf r},\omega)$ at the
next-nearest neighbor site of the Kondo hole, ${\bf R}^\prime$,
with $\Delta s({\bf R}^\prime)>0$, exhibits an increase in the
spectral weight at negative frequencies, and hence confirms this
conclusion (see Fig.~\ref{fig:spec}a).

The spatial perturbations in the magnetic correlations, $\chi({\bf
r, r'})$, possess a spectroscopic signature in the non-local
$f$-electron spin susceptibility, $\Gamma_f({\bf r, r'}, \omega)$,
between nearest neighbor sites ${\bf r}$ and ${\bf r'}$.
$\Gamma_f$ can in general be measured via atomic force microscopy
\cite{Ham04}, and its Fourier transform describes how magnetic
correlations decay in real time. A plot of Im$\Gamma_f({\bf r,
r'}, \omega)$ in Fig.~\ref{fig:spec}b illustrates the important
relation with $\Delta \chi$. In comparison to the unperturbed
Kondo lattice, one finds that for two sites ${\bf r},{\bf r'}$
with $\Delta \chi >0$, $|{\rm Im}\Gamma_f|$ is enhanced, while
$\Delta \chi <0$ leads to its suppression.

\begin{figure}[!t]
\includegraphics[width=8.5cm]{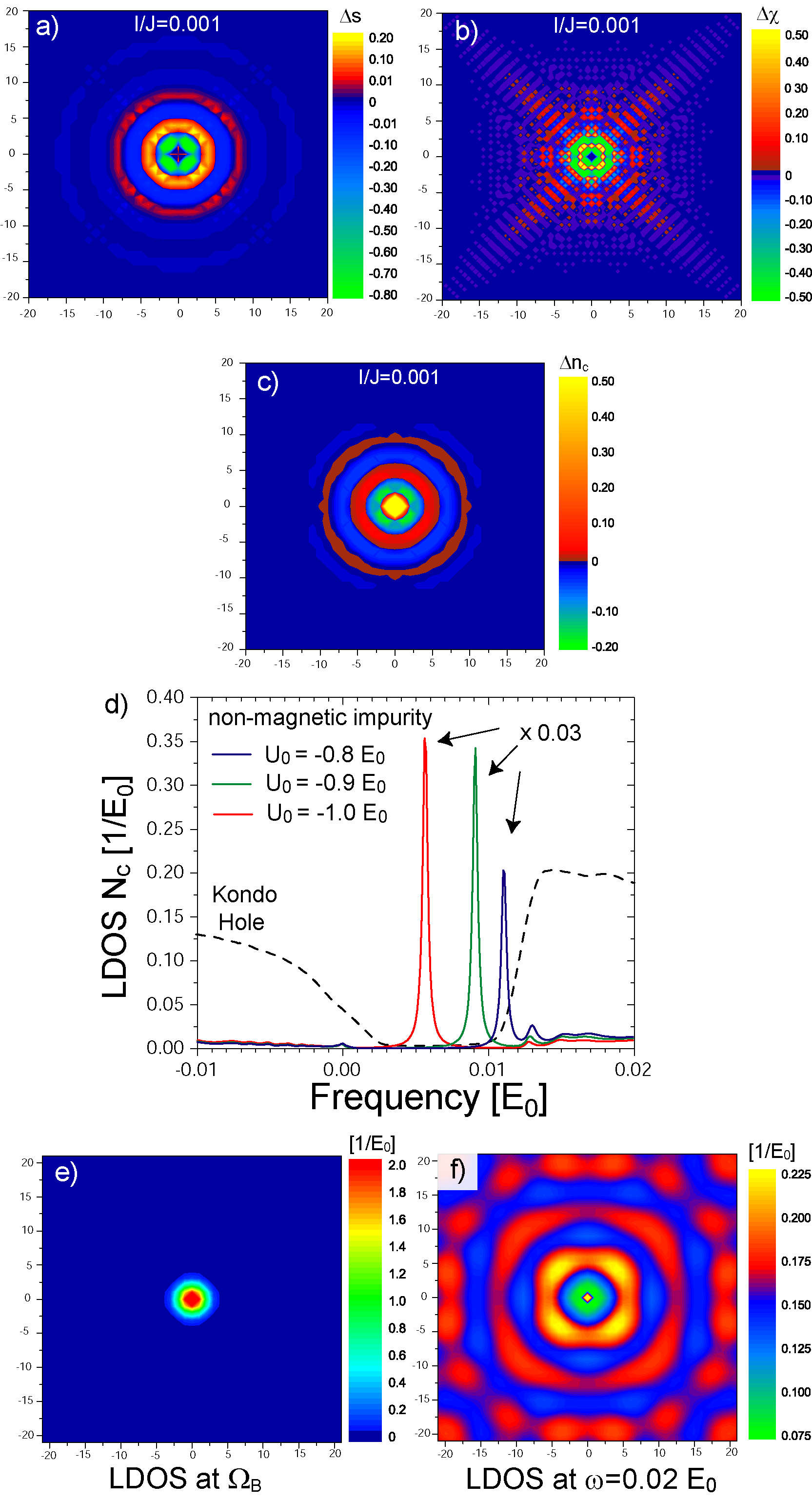}
\caption{{\bf Local density of states around a defect} Spatial
contour plot of {\bf a.} $\Delta s$, {\bf b.} $\Delta \chi$, and
{\bf c.} $\Delta n_c$ for a non-magnetic impurity with $U_0=-1.0
E_0$ and $I/J=0.001$. {\bf d.} Comparison of $N_c({\bf r},\omega)$
at the site of a Kondo hole and of a non-magnetic impurity. An
impurity bound state inside the hybridization gap emerges only for
an attractive non-magnetic impurity if $|U_0|>U_c=0.62 E_0$. To
facilitate the comparison, the curves of the bound state have been
multiplied by $0.03$. Spatial contour plot of $N_c({\bf
r},\omega)$ for $U_0=-1.0 E_0$ at {\bf e.} the energy of the bound
state, $\omega=\Omega_B$, and {\bf f.} outside the hybridization
gap at $\omega=0.02 E_0$.} \label{fig:Fig7}
\end{figure}

Replacing a Kondo atom by a non-magnetic impurity
\cite{Ste87,Lin87,Lop92,Law96} gives rise to a local scattering
potential for the conduction electrons (see Sec.~S2 of the
Supplemental Information). This, in turn, induces perturbations in
$\Delta n_c$, $\Delta s$, and $\Delta \chi$ (see
Figs.~\ref{fig:Fig7}a-c) that possess several distinct differences
to those caused by a Kondo hole. In particular, if the
non-magnetic scattering potential is attractive, $U_0<0$, the
spatial pattern of $\Delta n_c$ is inverted, i.e., a site with
$\Delta n_c >0$ for the Kondo hole case, now has $\Delta n_c <0$
(cf. Figs.~\ref{fig:Fig1}d and \ref{fig:Fig7}a). Since the same
inversion occurs for the spatial patterns of $\Delta s$ and
$\Delta \chi$, we conclude that the electronic and magnetic
correlations are strongly affected by the spatial redistribution
of $n_c$. Moreover, an {\it impurity bound state} can be formed
around the non-magnetic impurity, whose spectroscopic signature is
a sharp peak in $N_c({\bf r},\omega)$ inside the hybridization
gap, as shown in Fig.~\ref{fig:Fig7}d. The emergence of this
impurity state is tied to an attractive scattering potential,
$U_0<0$, whose magnitude exceeds a critical value, $U_c$, which is
determined by the particle-hole asymmetry of the conduction band.
A non-zero $U_c$ could be responsible for the observation that
different non-magnetic impurities can lead to disparate properties
of the system \cite{Ste87}. With increasing $|U_0|$, the bound
state first emerges at the high energy side of the hybridization
gap and then moves to lower energies, as shown in
Fig.~\ref{fig:Fig7}d. This bound state is predominantly formed by
$f$-electron states, as can be deduced from a spatial contour plot
of $N_c({\bf r},\omega)$ at the bound state energy, $\Omega_B$,
shown in Fig.~\ref{fig:Fig7}e. The bound state is spatially
isotropic, and decays exponentially with distance from the
non-magnetic impurity with a decay length, $\xi_D \approx 0.65
a_0$. If this bound state were formed by (light) conduction
electron states, a simple estimate shows that it would possess a
decay length in excess of $60 a_0$, in disagreement with the
results of Fig.~\ref{fig:Fig7}e. This result is particularly
interesting since the non-magnetic impurity leads to scattering in
the conduction band only. In contrast, the spatial oscillations of
$N_c({\bf r},\omega)$ for frequencies outside the hybridization
gap are delocalized, as exemplified in Fig.~\ref{fig:Fig7}f, and
hence are predominantly arising from the (light) conduction
electron band.

\begin{figure}[h]
\includegraphics[width=8.5cm]{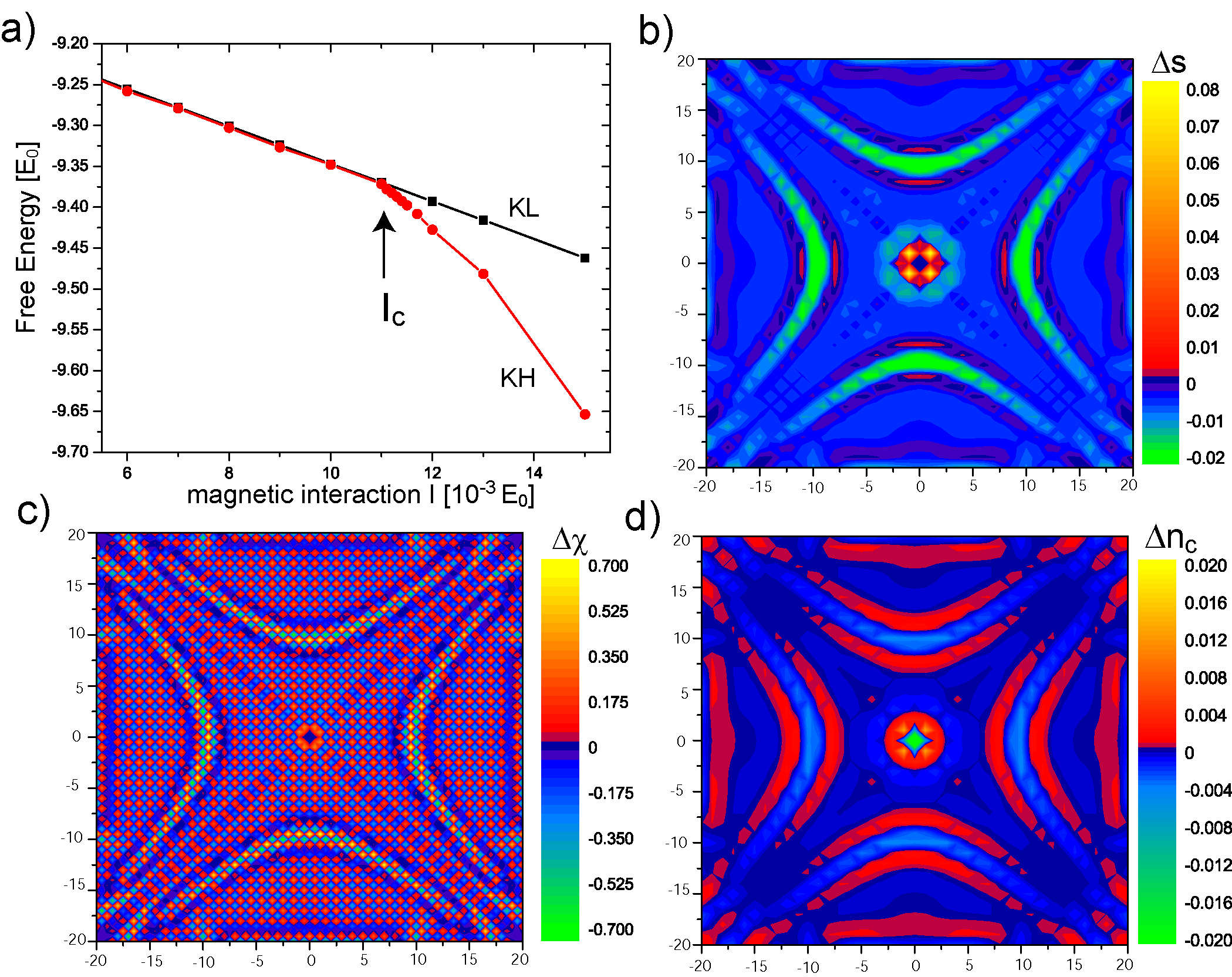}
\caption{{\bf Novel Ground State driven by Quantum Interference}
Periodic array of Kondo holes: {\bf a.} Free energy for an
unperturbed Kondo lattice (KL) and a Kondo hole (KH) array as a
function of $I/J$. For the electronic band structure considered
here, one has $I_c/J = 0.0011(1)$, indicated by an arrow. Contour
plots of {\bf b.} $\Delta s$, {\bf c.} $\Delta \chi$, and ({\bf
d.} $\Delta n_c$ for $I/J=0.0013>I_c/J$.} \label{fig:Fig3}
\end{figure}
The strongly correlated nature of the Kondo lattice also manifests
itself in highly non-linear quantum interference of the spatial
perturbations in $\chi$ and $s$ emanating from adjacent Kondo
holes. In a periodic array of Kondo holes, this non-linearity
drives the system to a novel inhomogeneous ground state via a
first order phase transition once $I$ exceeds a critical value,
$I_c$, as demonstrated by a plot of the free energy in
Fig.~\ref{fig:Fig3}a. Note that $I_c$ increases with increasing
distance between the Kondo holes. This novel inhomogeneous ground
state is characterized by spatial patterns of $\Delta \chi$,
$\Delta s$, and $\Delta n_c$ [see Figs.~\ref{fig:Fig3}b-d], that
are strikingly different from the ones for $I<I_c$ shown in
Fig.~\ref{fig:Fig1}. The similarity of these three spatial
patterns, as well as the significant amplitude of the
perturbations in the conduction electron density, $n_c$, suggest
that while the phase transition is driven by quantum interference
of the correlations' spatial perturbations, the resulting real
space patterns are determined by the redistributed conduction
electron density. Changing the symmetry of periodic array, for
example, by adding a second hole to the unit cell (see Fig.~S3 of
the Supplemental Information), leads to a significant change in
the
spatial pattern for $I>I_c$, reflecting the strong non-linearity of the system.    \\

%{\bf Acknowledgements}\\

We would like to thank E. Abrahams, P. Coleman, J.C. Davis, H.
Manoharan, S. Sachdev, Q. Si, F. Steglich and H. v. Lohneysen for
stimulating discussions. D.K.M. would like to thank the Aspen Center
for Physics and the James Franck Institute at the University of
Chicago for its hospitality during various stages of this project.
This work
is supported by the U.S. Department of Energy under Award No. DE-FG02-05ER46225.\\

\newpage

\begin{widetext}

\begin{center}
{\Large {\bf  Supplemental Online Information for} \\[0.5cm]
{\it Defects in Heavy-Fermion Materials:\\ Unveiling Strong
Correlations
in Real Space }\\[0.5cm]
by Jeremy Figgins and Dirk K. Morr}\\[1.cm]
\end{center}

\noindent {\bf Section S1: Supplemental Figures }\\

\setcounter{figure}{0}

\begin{figure}[h]
\begin{center}
\includegraphics[width=12.5cm]{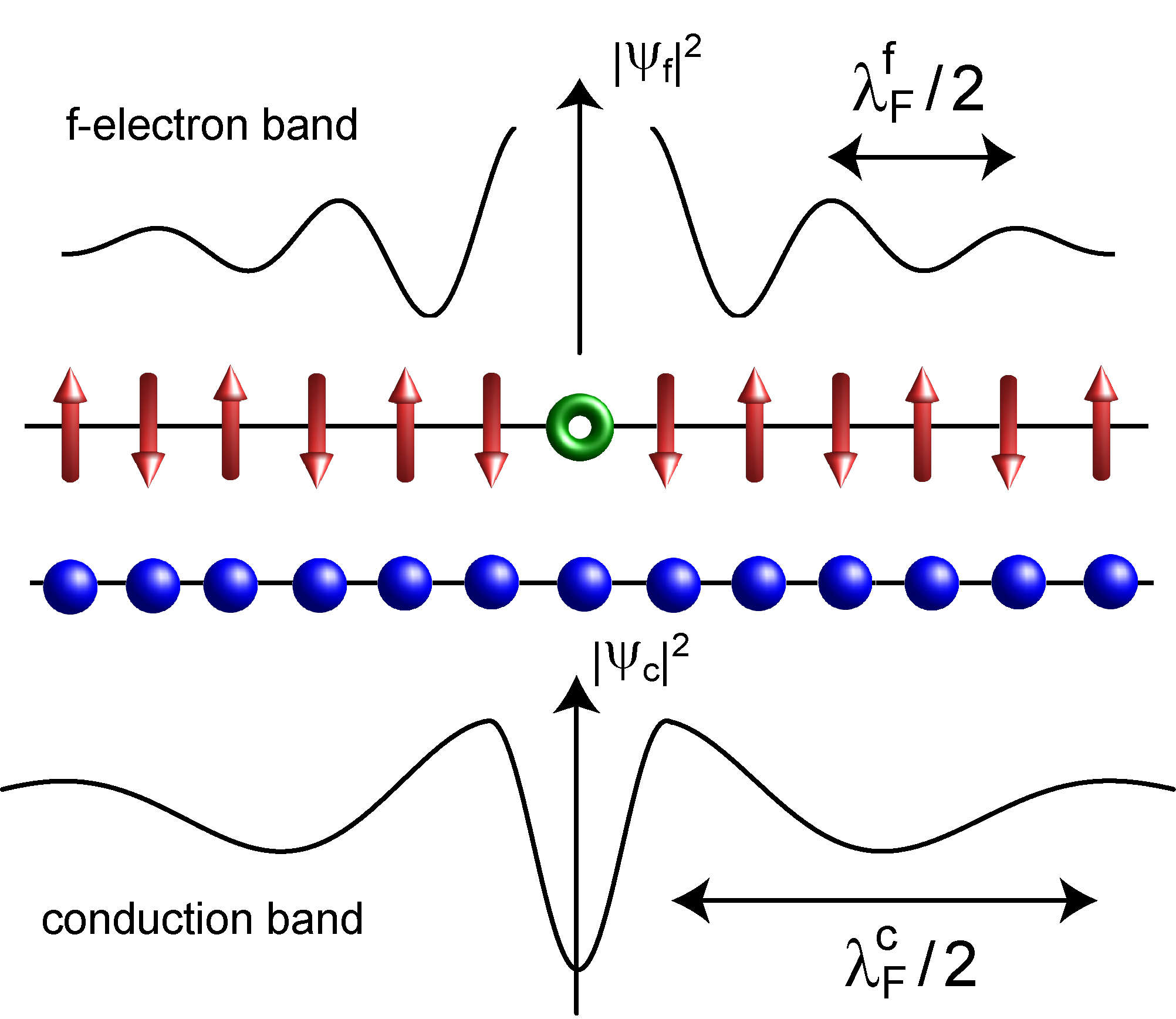}
\end{center}
\caption{{\bf Defect in a Kondo lattice} Blue dots represent the
sites of the metallic conduction lattice, while red arrows show the
locations of the magnetic moments, i.e., the magnetic Kondo atoms,
in the Kondo lattice. The green ring denotes the site of a missing
Kondo atom, i.e., a Kondo hole, or of a non-magnetic impurity, where
the latter leads to scattering of the conduction electrons only. The
defect induces spatial perturbations in the electronic and magnetic
correlations of the system, which are determined by the real space
oscillations of $|\Psi_c|^2$ with wavelength $\lambda_F^c/2$ and
$|\Psi_f|^2$ with wavelength $\lambda_F^f/2$. Here, $\Psi_c$ and
$\Psi_f$ are the wave functions of the conduction and $f$-electron
states, respectively, with $\lambda_F^c$ and $\lambda_F^f$ being
their associated Fermi wavelengths.} \label{fig:SupFig1}
\end{figure}

\newpage

\begin{figure}[!h]
\begin{center}
\includegraphics[width=10.cm]{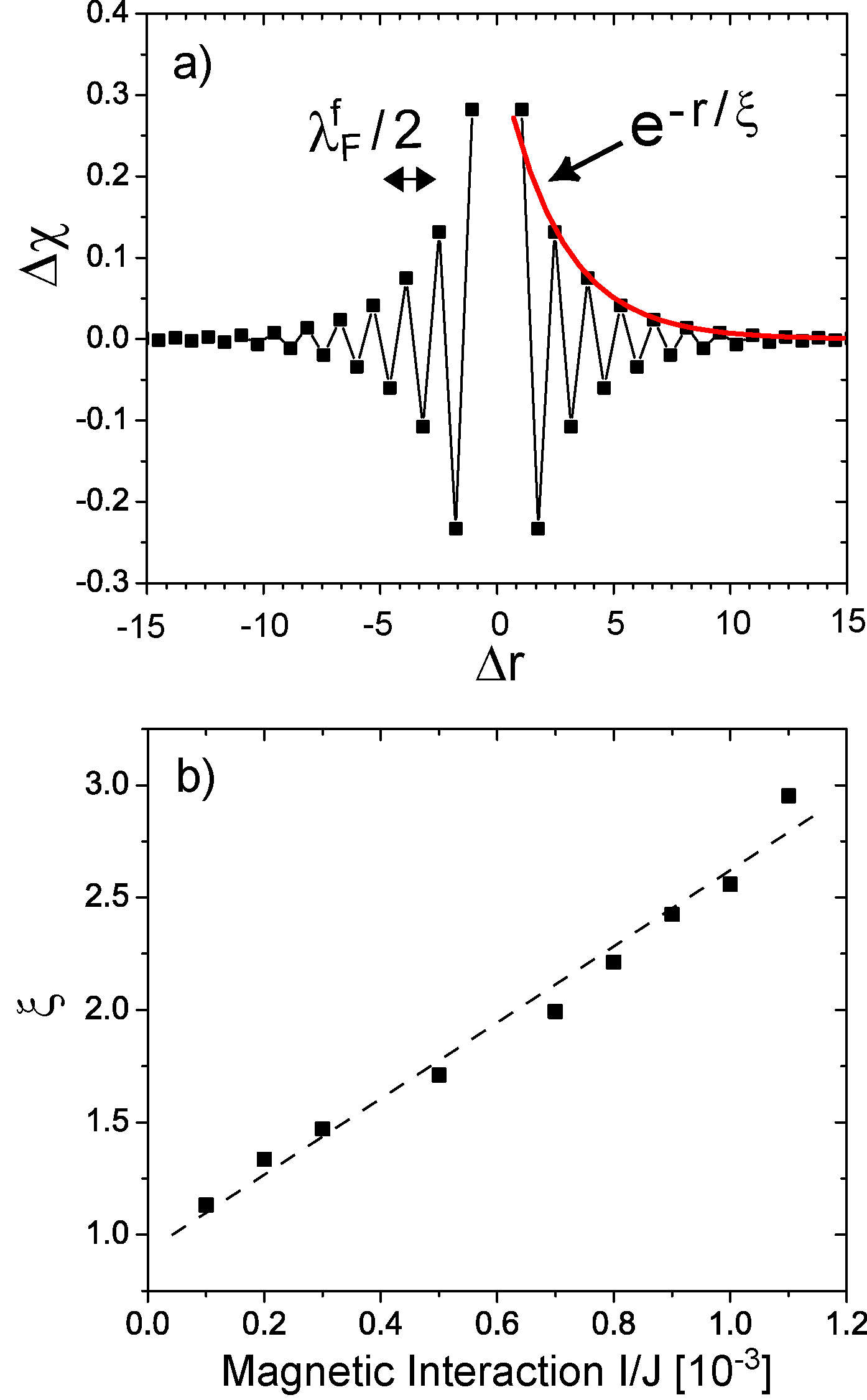}
\end{center}
\caption{ {\bf Spacial oscillations of $\Delta \chi$} a) Spatial
oscillations of $\Delta \chi({\bf r},{\bf r}')$ (shown as a function
of $\Delta r = |{\bf r} + {\bf r}'|/2$ measured from the Kondo hole)
along the lattice diagonal. The oscillations exhibit a wavelength
$\lambda_F^f/2$ and decay exponentially as $e^{-r/\xi}$ (the
exponential fit is shown as a red line). b) Dependence of $\xi$ on
the magnetic interaction, $I$. Errors in $\xi$ are smaller than the
data points. $\xi$ increases approximately linearly with $I/J$
(dashed line is a guide to the eye). }\label{fig:SupFig2}
\end{figure}
\newpage

\begin{figure}[!h]
\includegraphics[width=10.cm]{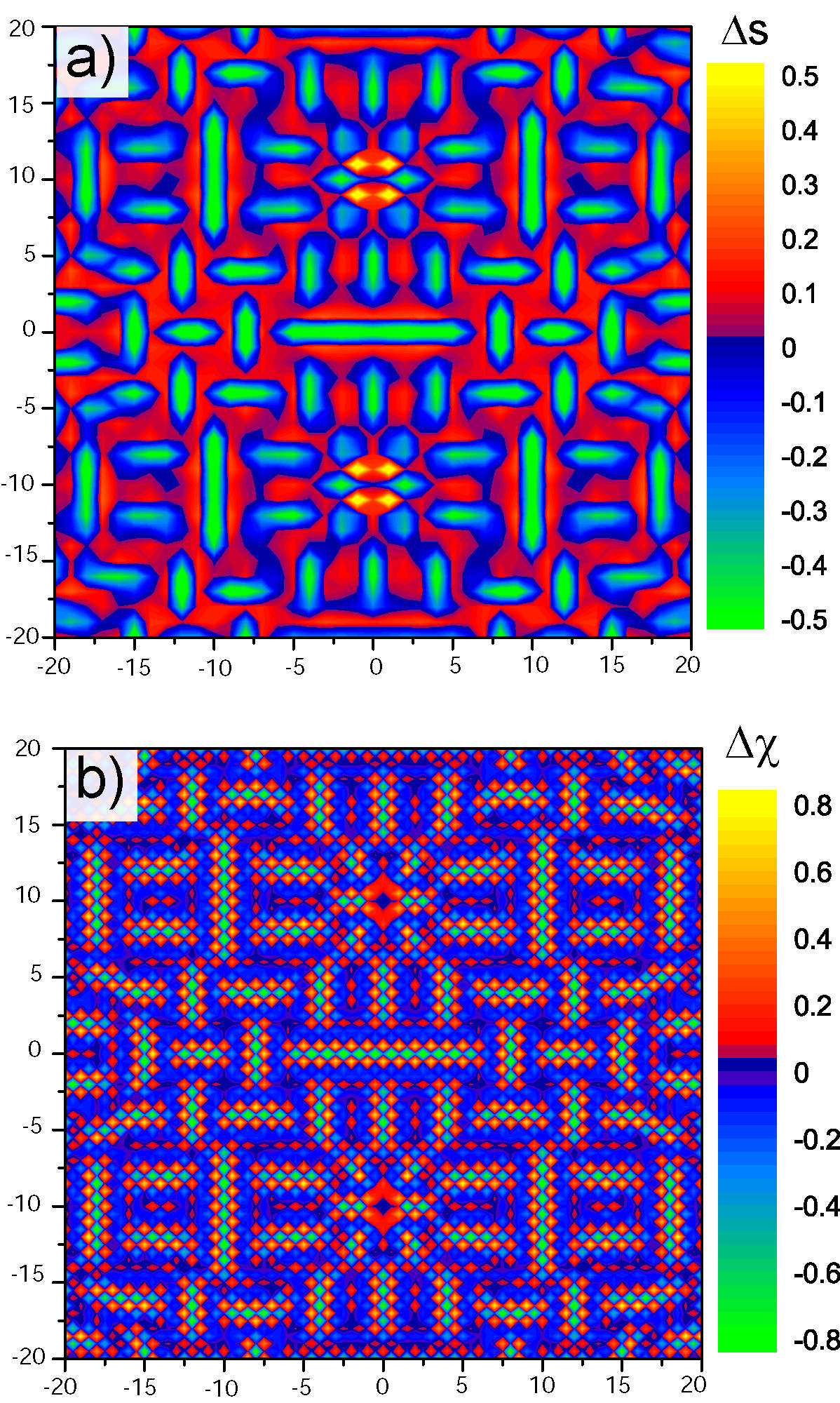}
\caption{{\bf Quantum Interference in a periodic array of Kondo
holes with two Kondo holes per unit cell} Contour plot of (a)
$\Delta s$ and (b) $\Delta \chi$ for $I/J=0.0013>I_c/J$. The spatial
interference pattern for $I>I_c$ is distinctively different from
that of a periodic array with a single Kondo hole per unit cell
(shown in Fig.~4), reflecting the strong non-linearity of the
system.} \label{fig:2KH}
\end{figure}

\newpage

\noindent {\bf Section S2: The Kondo-Bogoliubov-de Gennes (KBdG) Formalism}\\

Starting point for the study of defects in Kondo lattice systems is
the Kondo Heisenberg Hamiltonian
\begin{equation}
{\cal H} = -\sum_{{\bf r,r'},\sigma} t_{{\bf r,r'}} c^\dagger_{{\bf
r},\sigma} c_{{\bf r'},\sigma} + J {\sum_{{\bf r}}}' {\bf
S}^{K}_{\bf {\bf r}} \cdot {\bf s}^c_{\bf r} + {\sum_{{\bf r,r'}}}'
I_{{\bf r,r'}} {\bf S}^{K}_{\bf r} \cdot {\bf S}^{K}_{\bf r'}
\tag{S1} \label{eq:1} \ ,
\end{equation}
where $t_{{\bf r,r'}}$ is the fermionic hopping element between
sites ${\bf r}$ and ${\bf r'}$, and $c^\dagger_{{\bf r},\sigma},
c_{{\bf r},\sigma}$ creates (annihilates) a conduction electron with
spin $\sigma$ at site ${\bf r}$. $J>0$ is the Kondo coupling, and
${\bf S}^{K}_{\bf r}$ and ${\bf s}^c_{\bf r}$ are the $S=1/2$ spin
operators of the magnetic (Kondo) atom and the conduction electron
at site ${\bf r}$, respectively. $I_{{\bf r,r'}}>0$ is the
antiferromagnetic interaction between Kondo atoms at sites ${\bf r}$
and ${\bf r'}$, which we restrict to nearest-neighbor sites below.
We take the lattices of the conduction band and that of the magnetic
(Kondo) atoms to possess identical structures and lattice constants.
Thus, the site of a Kondo atom and the (single) conduction site that
it couples to are denoted by the same ${\bf r}$. The unprimed sum
runs over all sites of the conduction lattice, while the primed sum
runs over all positions of the Kondo atoms. These two sums are
identical for an unperturbed Kondo lattice without any defects.

A systematic large-$N$ expansion
\cite{Col83,Read83,Mil87,Bic87,Aff88,Hew93a,Sen04,Kaul07,Paul07a} of
the Hamiltonian can be achieved by generalizing the spin operators
to $SU(N)$ and representing them using Abrikosov pseudofermions
\begin{equation}
{\bf S}^{K}_{\bf r} =  \sum_{\alpha, \beta} f^\dagger_{{\bf
r},\alpha} {\bm \sigma}_{\alpha, \beta} f_{{\bf r},\beta} \quad {\bf
s}^{c}_{\bf r} =  \sum_{\alpha, \beta} c^\dagger_{{\bf r},\alpha}
{\bm \sigma}_{\alpha, \beta} c_{{\bf r},\beta} \ , \tag{S2}
\label{eq:slave}
\end{equation}
where $\alpha,\beta =1,..., N$ and ${\bm \sigma}_{\alpha, \beta}$
are the generators of $SU(N)$. Here, $f^\dagger_{{\bf r},\alpha}$
($f_{{\bf r},\alpha}$) creates a pseudofermion in the $f$-electron
state of the magnetic Kondo atom. In addition, one needs to satisfy
the constraint that each magnetic atom (i.e., each $f$-electron
site) is occupied by $N/2$ particles, i.e.,
\begin{equation}
{\hat n}_f({\bf r}) = \sum_\alpha f^\dagger_{{\bf r},\alpha} f_{{\bf
r},\alpha} = \frac{N}{2} \ . \tag{S3}
\end{equation}
Inserting the representations of Eq.(\ref{eq:slave}) into the
Hamiltonian,Eq.(\ref{eq:1}), yields quartic fermionic interaction
terms. Since our objective is the study of Kondo lattices containing
defects which break the translational invariance, it is necessary to
introduce a mean-field decoupling of the resulting interaction terms
in real space. The mean-field theory of the homogeneous Kondo
lattice (or the single Kondo impurity) becomes exact in the limit $N
\rightarrow \infty$ \cite{Col83,Read83}. However, it was argued that
even for $N=2$, the qualitative and to a large extent quantitative
features of the mean-field solutions remain unchanged
\cite{Bic87,Paul07a}, which in certain limits has been validated by
the comparison with exact numerical renormalization group studies
\cite{Sil96}. Following this argument, we set $N=2$ below, and
introduce the following local and non-local expectation values
\begin{equation}
s({\bf r}) = \frac{J}{2}\sum_{\alpha} \langle f^\dagger_{{\bf
r},\alpha} c_{{\bf r},\alpha} \rangle \, ; \quad \chi({{\bf r,r'}})
= \frac{I_{\bf r,r'}}{2}\sum_{\alpha} \langle f^\dagger_{{\bf
r},\alpha} f_{{\bf r'},\alpha} \rangle \label{eq:2} \ . \tag{S4}
\end{equation}
Here, $s({\bf r})$ describes the local hybridization between the
conduction electron states and the magnetic $f$-electron states,
whose magnitude is a measure of the Kondo screening. Thus $s({\bf
r})=0$ represents an unscreened magnetic moment at site ${\bf r}$.
$\chi({{\bf r,r'}})$ is a measure of the magnetic correlations
between Kondo atoms \cite{Sen04,Paul07a}. To enforce the constraint
$\langle {\hat n}_f({\bf r}) \rangle = 1$ within our formalism, we
add the term $\sum_{{\bf r},\alpha} \varepsilon_f({\bf r})
f^\dagger_{{\bf r},\alpha} f_{{\bf r},\alpha}$ to the Hamiltonian in
Eq.(\ref{eq:1}), where $\varepsilon_f({\bf r})$ represents the
on-site energy of the $f$-electrons. The resulting Hamiltonian is
quadratic, containing the set of parameters $\{ s({\bf
r}),\chi({{\bf r,r'}}), \varepsilon_f({\bf r}) \}$, and therefore
can be diagonalized in real space (we assume periodic boundary
conditions). After each diagonalization, the mean-fields $s({\bf
r})$ and $\chi({{\bf r,r'}})$ are computed self-consistently via
Eq.(\ref{eq:2}), and $\varepsilon_f({\bf r})$ is chosen such that
$\langle {\hat n}_f({\bf r}) \rangle=1$ at each site. This procedure
is repeated until a self-consistent solution for $\{ s({\bf
r}),\chi({{\bf r,r'}}), \varepsilon_f({\bf r}) \}$ is obtained. We
note that for an unperturbed Kondo lattice, this formalism is
identical to the saddle-point approximation of the
path-integral approach \cite{Col83}. \\

In order to describe a Kondo hole at site ${\bf R}$, we remove the
corresponding spin operator, ${\bf S}^{K}_{\bf R}$ from the
Hamiltonian in Eq.(\ref{eq:1}) \cite{Kaul07}. When a Kondo atom at
site ${\bf R}$ is replaced by a non-magnetic impurity, the spin
operator ${\bf S}^{K}_{\bf R}$ is removed, and the term $U_0
\sum_\alpha c^\dagger_{{\bf R},\alpha} c_{{\bf R},\alpha}$ is added
to the Hamiltonian, where $U_0$ is the non-magnetic scattering
strength. Here, we assume that the non-magnetic atom leads to
scattering of the conduction electrons only, since the magnetic
$f$-electron states are not expected to hybridize with the
electronic states of the non-magnetic impurity, in agreement
with Ref.~\onlinecite{Fre90}, but in contrast to Ref.~\onlinecite{Sol91}.\\

For the results shown in the article, we took the conduction and
Kondo lattices to be finite two-dimensional square lattices
containing each ${\cal N}=M \times M$ sites. Our formalism can
equally well be applied to three-dimensional lattices, however, we
do not expect that the qualitative nature of the results shown in
this article will be changed. For the conduction band, we employ a
nearest-neighbor hopping, $t=0.5 E_0$, and a chemical potential,
$\mu=-1.809 E_0$. Here, $E_0$ is an overall energy scale related to
the conduction bandwidth $W=4E_0$. These parameters yield an
isotropic conduction Fermi surface (see Fig.~1b) with
Fermi-wave-length $\lambda^c_F=10 a_0$, where $a_0$ is the lattice
constant, and a band filling (per spin degree) of $\pi/100$.\\

We choose the size of the lattice sufficiently large such that the
set of solutions for $\{ s({\bf r}),\chi({{\bf r,r'}}),
\varepsilon_f({\bf r}) \}$ is independent of ${\cal N}$. The results
shown in the main text were obtained for systems with $M=41$, and we
did not find any significant quantitative changes for lattice sizes
up to $M=71$. For the unperturbed Kondo lattices with $J=1.0 E_0$,
the set of parameters for several values of $I/J$ are shown in Table
1.\\

\begin{table*}[h]
 \label{tab:1}
  \begin{center}
    \begin{tabular}{ | p{2.0cm} || p{2cm} | p{2cm} | p{2cm} |}
    \hline
      \begin{center} $I/J$ \end{center} & \begin{center} $\varepsilon_f$ \end{center}  & \begin{center} $s$ \end{center} & \begin{center} $\chi$ \end{center}
      \\
    \hline
    \hline
     0.0005
     &  0.001247
     &  0.048470
     &  0.00008122
     \\
    \hline
     0.001
     & 0.001232
     & 0.048470
     & 0.00016579 \\
     \hline
     0.0011
     & 0.001229
     & 0.048470
     & 0.00018281 \\
     \hline
     0.0013
     & 0.001225
     & 0.048470
     & 0.00021688 \\
    \hline
    \end{tabular}
\end{center}
 \caption{Values for $s, \chi$, and
$\varepsilon_f$ (in units of $E_0$) for an unperturbed Kondo lattice
as a function of $I/J$. Note that for these small values of $I/J$,
$s$ remains practically unchanged. A significant reduction of $s$ is
only seen for larger values of $I/J$.}
\end{table*}

\noindent {\bf Section S3: The Local Density of States of the Conduction Band}\\[0.1cm]

In general, the local density of state for the conduction band,
$N_c({\bf r},\omega)$, can be directly obtained from the
eigenvectors and energies of the diagonalization procedure described
in the previous section. However, the finite size of the systems
leads to (artifactual) strong oscillations in the frequency
dependence of the LDOS. We developed an alternative approach to
circumvent this problem, and verified that it significantly reduces
the artifactual frequency oscillations, and that it yields identical
results to the first approach when considering the spatial form of
$N_c({\bf r},\omega)$ at fixed $\omega$. In this approach, we
compute the local conduction electron density of states at a site
{\bf r} for a given set of solutions, $\{s({\bf r}), \chi({{\bf
r,r'}}), \varepsilon_f({\bf r}) \}$ via
\begin{equation}
N_c({\bf r},\omega) = -{1 \over \pi} {\rm Im} [G({\bf r,r},\omega)]
\tag{S5} \ ,
\end{equation}
where $G({\bf r,r},\omega)$ is the full retarded conduction band
Green's function given by (in what follows, all Greens functions are
retarded)
\begin{equation}
G({\bf r,r},\omega) = G_0({\bf r,r},\omega) + \hat{G}^{(1)}_0(\Delta
{\bf r}, \omega) \hat{s} \hat{F}( \omega) \hat{s}
\hat{G}^{(2)}_0(-\Delta {\bf r}, \omega) \label{eq:fullG} \ .
\tag{S6}
\end{equation}
Here we defined
\begin{equation} \hat{G}^{(1)}_0(\Delta {\bf r}, \omega) = \left( G_0( {\bf r},
{\bf r}_{1}, \omega), ... , G_0({\bf r}, {\bf r}_{{\cal P}},
\omega)\right) \ ; \quad \hat{G}^{(2)}_0(-\Delta {\bf r}, \omega) =
\left(
\begin{array}{c} G_0( {\bf r}_{1},
{\bf r}, \omega) \\
\vdots \\
G_0({\bf r}_{{\cal P}},{\bf r},  \omega)
\end{array}
\right) \tag{S7} \ ,
\end{equation}
where $G_0$ is the Greens function of the decoupled (unhybridized)
conduction band, ${\bf r}_i$ is the position of the $i'th$ Kondo
atom, and ${\cal P}$ is the number of Kondo atoms in the system.
Moreover, for the hybridization matrix $\hat{s}$ one has
\begin{equation}
\hat{s} = \left(
\begin{array}{ccccc}
s({\bf r}_1) & 0 & 0 & 0 & 0 \\
0 & s({\bf r}_2) & 0 & 0 & 0 \\
0 & 0 & ... & 0 & 0 \\
0 & 0 & 0 & s({\bf r}_{{\cal P}-1}) & 0 \\
0 & 0 & 0 & 0 & s({\bf r}_{{\cal P}})%
\end{array}
\right) \tag{S8}  \ , \end{equation}  and $\hat{F}$ is the full
$f$-electron Green's function with the $(ij)$ element of its inverse
given by
\begin{equation}
\left[ \hat{F}(\omega)^{-1}\right] _{ij}=\left\{
\begin{array}{c}
\omega-\varepsilon_{f}({\bf r}_{i})-s({\bf r}_{i})G_{0}({\bf
r}_{i},{\bf r}_{j},\omega)s({\bf r}_{j})
\text{ \ \ \ \ \ \ \ \ if }i=j \\
\chi({\bf r}_i,{\bf r}_j)-s({\bf r}_{i})G_{0}({\bf r}_{i},{\bf
r}_{j},\omega)s({\bf r}_{j}) \text{ \ \ \ \ \ \ \ \ \ \ \ \ \ if
}i\neq j \ .
\end{array}
\right. \label{fullF} \tag{S9}
\end{equation}

In the presence of a non-magnetic impurity scatterer at site ${\bf
R}$ with potential $U_0$, it is easy to show that the full
conduction electron Green's function can be obtained from
\begin{equation}
G_{NM}({\bf r,r},\omega) = G({\bf r,r},\omega) + \frac{ G({\bf
r,R},\omega) U_0 G({\bf R,r},\omega) }{ 1 - U_0 G({\bf R,R},\omega)
} \tag{S10}
\end{equation}
with $G({\bf r,r'},\omega)$ obtained from Eq.(\ref{eq:fullG}).\\

\noindent {\bf Section S4: The $f$-electron Spin Susceptibility}\\[0.1cm]

The $f$-electron spin susceptibility, $\Gamma_f$, in Matsubara time
is given by
\begin{equation}
\Gamma_f({\bf r}, {\bf r}', \tau) =  \langle T_\tau {\bf S}^{K}_{\bf
{\bf r}}(\tau) \cdot {\bf S}^{K}_{\bf {\bf r}'}(0) \rangle \ .
\tag{S11}
\end{equation}
After Fourier transformation into Matsubara frequency space, and
analytic continuation, one obtains ($k_B=1$)
\begin{equation}
\Gamma_f({\bf r}, {\bf r}', i \Omega_m) =  \frac{3}{2} T \sum_n
F({\bf r}, {\bf r}', i \omega_n) F({\bf r}',{\bf r},  i \omega_n+i
\Omega_m) \ , \tag{S12}
\end{equation}
where $F$ is the $f$-electron Greens functions from
Eq.(\ref{fullF}). Using next the Lehmann representation of $F$, one
obtains for the imaginary part of the retarded $f$-electron spin
susceptibility
\begin{equation}
{\rm Im}\, \Gamma_f({\bf r ,r'},\Omega )=\frac{3}{2\pi
}\int_{0}^{\Omega }d\upsilon \, {\rm Im}F({\bf r,r'},\upsilon )\,
{\rm Im} F({\bf r',r},\upsilon -\Omega )  \ . \tag{S13}
\end{equation}

\end{widetext}

\end{document}